# Feasibility study of multiplexing analog signals from SiPMs for a single layer monolithic PET detector design

**Shiv K. Subedi[1*], Simon R. Cherry[2], Yi Qiang[1], Peng Peng[1*]**

[1]Canon Medical Research USA, Inc., Vernon Hills, IL 60061, USA

[2]Department of Biomedical Engineering, University of California, Davis, CA 95616, USA

**Email:** ssubedi@mru.medical.canon, ppeng@mru.medical.canon

**Abstract**

Semi monolithic detector designs with a series of stacked thin monolithic scintillator plates and side readout are an attractive approach for potentially achieving very high performance in a positron emission tomography (PET) scanner. In this work, a simulation study of a single layer monolithic detector module was performed with side read out of scintillation light using GATEv8.2. In this design, a single layer LSO crystal was used with dimensions **40 mm × 40 mm × 4 mm**, with 0.60 mm thickness of the ESR (enhanced specular reflector) films covering the crystal's top and bottom surfaces. The photons generated in the scintillation process induced by the gamma ray hitting the crystal were detected by four 1×8 SiPM (silicon photomultiplier) arrays placed along the four sides of the crystal. The scintillation light distribution detected by all of the 32 SiPMs surrounding the crystal layer was then used to extract the gamma-crystal interaction location based on machine learning analysis. In this work, the spatial resolution of the detector module was explored when analog signals from each of the 32 SiPMs were summed to 28, 24, 20, 16, 12, 8, and 4 total outputs. This study showed that good spatial resolution can be achieved even when the number of read out channels is decreased by multiplexing, which can reduce the overall detector manufacturing cost.

(Some figures may appear in colour only in the online journal)

## 1. INTRODUCTION

Previous work from the PET community has shown that the semi-monolithic design of scintillation crystals in a detector module can potentially provide better performance (resolution and sensitivity) over a traditional pixelated design [1] [2] [3] [4] [5] [6] [7] [8]. This work adopted a layered monolithic crystal design for detector geometry and implemented a side read out using silicon photomultipliers (SiPMs) placed on all four sides of the crystal. Compared to long narrow pixelated crystals which exhibit significant loss of scintillation light as photons are transported to the photosensor through many reflections. The four-sided readout design reduces the number of reflections during light transport, hence improves the light collection efficiency, and in turn improves the energy and timing resolution [9].

There are several other advantages with the semi-monolithic crystal design. This geometry provides an advantage of extracting depth of interaction (DOI) information when multiple layers of crystal are stacked, which opens the door for building smaller diameter PET detector (decreasing the manufacturing cost while increasing sensitivity) without degrading spatial resolution caused by parallax errors. In addition, compared to the pixelated crystal design which inserts reflective materials in between crystals, the semi-monolithic design has continuity of crystal material which increases sensitivity for gamma-ray detection. Machine learning techniques are used to decode the interaction position of the gamma-ray in each crystal layer based on the light distribution on the SiPMs which enabled sub-millimeter spatial resolution to be achieved [4] [5] (Note: sub-millimeter resolution has also been achieved using conventional pixelated detector designs, but to date only for small animal PET systems [10] [11] [12]). Also, the interaction location and energy resolution of each layer can be evaluated independently. This provides the possibility of stacking more crystal layers in a detector module to increase system sensitivity without affecting the spatial and energy resolution of the system.

There has been active research on studying the effect of output signal multiplexing on the spatial resolution. Previous work on multiplexing in PET detector development have adopted in general two methods: light multiplexing (i.e., light sharing) [13] and charge multiplexing [14] [15] [16] [17] [18] [19]. Irrespective of the method, the goal of such studies is to explore optimal multiplexing configuration of signals collected by photodetectors that can provide similar/better spatial resolution compared to no-multiplexing. If such an optimal multiplexing configuration can be found, it provides an opportunity to reduce the associated digitization cost of analog signals from photodetectors by reducing the number of acquisition channels and reducing power consumption as well.

In this work, several variations of multiplexing were studied in which different combinations (i.e., summing) of the analog signals from the SiPMs were explored before digitization. For this, first the number of optical photons collected by individual 32 SiPMs on four sides of the LSO crystal from scintillation process induced by gamma-rays hitting the crystal were recorded. Then, various multiplexing (i.e., summation) schemes for the SiPM signals were implemented. Though this method of multiplexing is different from past studies, there is a common objective of determining the influence of multiplexing on the spatial resolution of the detector module.

## 2. MATERIAL AND METHODS

### 2.1 Implementing GATE simulations

GATE v8.2 (**G**eant4 **A**pplication for **T**omographic **E**mission) [20] [21] [22] was used to design the detector geometry and perform optical simulations for generating the scintillation photons. The detector design comprised of an LSO crystal plate with dimensions $40\ mm \times 40\ mm \times 4\ mm$, with 0.60 mm-thick layers of ESR reflector on the top and bottom of the crystal surfaces. A total of 32 SiPMs were distributed along the four sides of the crystal with 8 SiPMs covering each side as shown in Figure 1. The SiPMs have dimensions of $4\ mm \times 4\ mm$ and were placed with a pitch of 4.6 mm along each side of the crystal. The optical photons collected by the individual 32 SiPMs were used to test the various read-out schemes. Since the light collection efficiency and distribution are independent of the number of layers for the semi-monolithic design, the single layer detector studied in this work is sufficient to evaluate the different read-out schemes.

In the GATE simulations, the following physics processes were included: photoelectric interaction, Compton scattering, Rayleigh scattering, Bremsstrahlung, electron ionization, electron multiple scattering, Mie scattering, optical scattering and absorption, optical boundaries, and scintillation. As seen from the detector geometry (Figure 1), there are two possible interface boundaries for scintillation photons as they propagate through the crystal's volume: 1. crystal and ESR interface; 2. crystal and SiPM interface. The boundary across crystal and ESR interface was defined as *by a 'dielectric-dielectric' type with* reflectance = 0.98. Similarly, the boundary between crystal and SiPM was defined by a '*dielectric-metal*' type with no reflectance and efficiency = 0.5. Both the boundary definitions above were implemented using the UNIFIED Model.

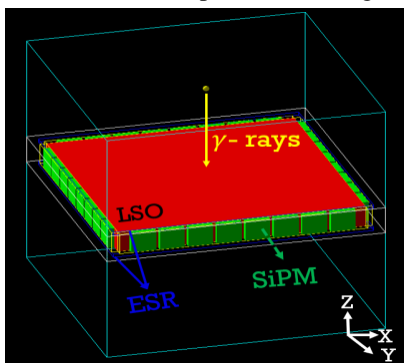

*Figure 1: Schematic diagram showing the geometry of single layer monolithic PET detector module adopted in this work. The LSO crystal block with dimensions of 40 mm×40 mm×4 mm is shown in red. On the top and bottom of the crystal block there is an ESR layer with thickness of 0.60 mm (shown in the figure in the form of a thin wire frame in blue). The SiPMs used in the design have dimensions of 4 × 4 mm and the pitch of the SiPM array is 4.6 mm, as shown in green.*

As shown in Figure 1, for the simulations, 511 keV monoenergetic gamma-rays emitted from a point source were incident normal to the top surface of the crystal, producing scintillation photons. The numbers of scintillation photons detected by each of the 32 SiPMs were then recorded, and later used to extract the interaction location of gamma-rays within the crystal based on Machine Learning (ML) analysis.

The objective of this work is to combine analog signals from the SiPMs through various multiplexing schemes before digitization and investigate the corresponding impact on the spatial resolution of the detector module. The dominant noise source is assumed to be the statistical uncertainty in the number of detected optical photons. In comparison, the electronic noise added by the SiPM is expected to be negligible based on specifications of latest generation SiPMs. Thus, electronic noise was not considered. The number of readout channels was reduced from 32 (no multiplexing) to 28, 24, 20, 16, 12, 8, and 4. For the 28, 24, and 20 readout channels, only one multiplexing scheme each was tested and for 16, 12, 8, and 4 readout channels, several possible multiplexing schemes were investigated. The details of each scheme are presented later.

### 2.2 Machine learning architecture and application

In order to extract the interaction location of gamma-rays, a convolutional neural network (CNN) architecture (using the Neural Network Toolbox in MATLAB R2021b) was used. For this, first, the scintillation light distribution collected from the SiPMs was converted into corresponding grayscale images. In Figure 2, sample grayscale images generated from a 32-channel readout configuration are shown.

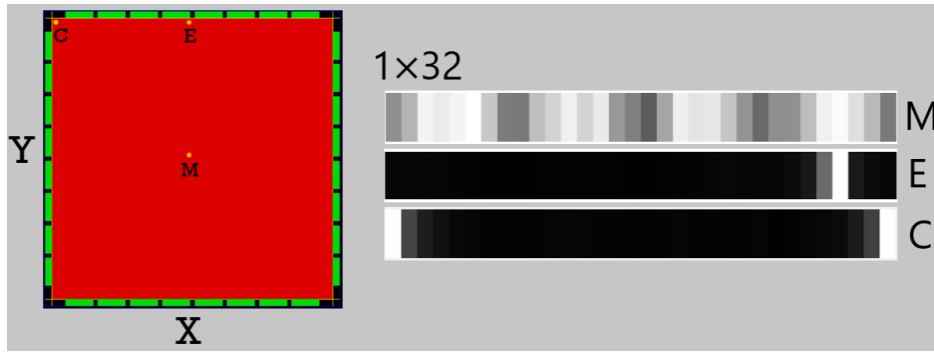

*Figure 2: Light distribution from individual 32 SiPMs (no multiplexing) for interaction positions near the middle (M), edge (E), and corner (C) regions of the detector plane converted into corresponding 1 × 32 grayscale images.*

Then, the CNN, with its structure described in Figure 3, was applied on the grayscale images generated from the SiPM signals and the output is the probability of the event occurring at each of the N calibration (training) points (for our case 'N' = 1600).

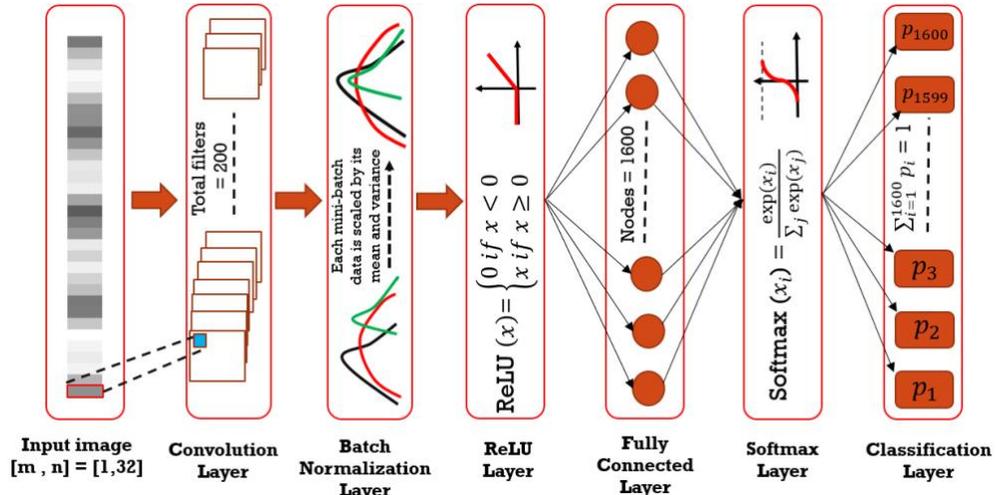

*Figure 3: CNN architecture implemented in this work for decoding the interaction position. Here, an example is presented where a 1 × 32 grayscale image is provided to the input layer. Based on the optimization test performed of number of filters w.r.t training accuracy and training time, 200 filters were used in the convolution layer. The convolution layer is specialized to extract the feature maps of the images from the input layer. 'N' number of nodes were provided to the fully connected layer, where 'N' refers to the corresponding gamma-source positions in the training data (for our case 'N' = 1600).*

These probabilities ($p_1, p_2, \ldots, p_{1600}$) were then used as weights in a center-of-mass calculation (Equation 1) to estimate the interaction position for the event i.e., [$x(predicted)$, $y(predicted)$].

$$x(predicted) = \frac{\sum_{i=1}^{N=1600} p_i \times x(calibration)_i}{\sum_{i=1}^{N=1600} p_i}, \qquad (1)$$

where the denominator sums to unity. Likewise, '$y(predicted)$' was also evaluated following Equation 1. The size of the network depends on the number of input channels and number of calibration points.

Before settling with these choices of CNN architecture and training options, an optimization study of the performance of the neural network was performed. For example, for optimization of the CNN, evaluated choices such as the use of multiple (1,2, and 3) layers were used, for the convolution number, the number of nodes were varied from low (100) to high (2000) and for node = 200, the filter size and stride were varied from 1-3 respectively, the position of reluLayer was switched before the convolution layer, batch normalization layer was removed, and a max pooling layer was added/removed. Likewise, the following choices for the training options were investigated: changing the optimization algorithm to 'Adam' and 'RMSProp', varying max epoch to 20, 25, 30, and varying the initial learning rate to 0.01 and 0.0001. However, these choices of CNN architecture and training options were either worse or only in some cases equal to the optimal result we obtained with the chosen architecture and parameters and didn't necessarily improve the relative performance. Hence, the CNN architecture as shown in Figure 3 was used in all studies.

The training data were generated by sweeping a point gamma-source with perpendicular emission across the top of the crystal's surface at 1600 different calibration positions (forming a 40 × 40 grid) along the X-Y plane with a pitch of 1 mm, as shown in Figure 4 (left). Training was performed with maximum training epoch = 10, initial learning rate = 0.001, using 90% of the total training dataset. Stochastic gradient descent with momentum (sgdm) was used as the optimization algorithm to converge to the global minima. The remaining 10% of the training dataset was used as validation dataset in tuning the model's hyperparameters.

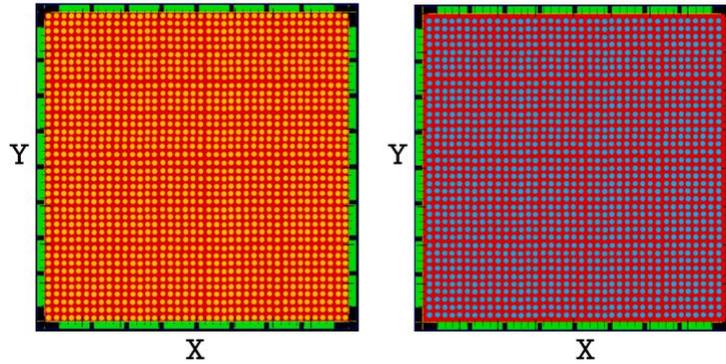

*Figure 4: Schematic diagrams presenting the gamma-source incident positions corresponding to our training (left) and testing dataset (right). (a) The training dataset was generated by sweeping gamma rays perpendicularly on a 40 × 40 grid with 1 mm pitch along X-Y plane at 1600 total positions. (b) The test dataset was generated in a similar fashion as training data, but here the gamma-source was swept at 1521 total positions placed in the center of four adjacent training positions forming a 39 × 39 grid with 1 mm pitch along the X-Y plane.*

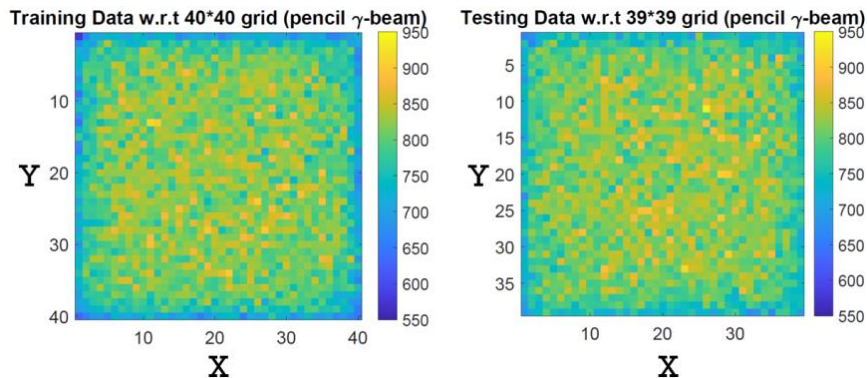

*Figure 5: Distribution of events i.e., total gamma-rays incident on each grid for training and test dataset.*

The testing dataset was generated using the same methodology as the training dataset, with the gamma-rays positioned at an offset of 0.5 mm (half of the pitch) in both X and Y directions compared to the grid of the training data. This thus formed a test dataset with a 39 × 39 grid and total of 1521 total gamma-ray positions, as shown in Figure 4 (right). For both data sets, simulations were performed with a low activity (10 becquerel) 511 keV monoenergetic gamma-ray source for 500 seconds at each position. In Figure 5, the number of events within the 20% energy window of 408.8 – 613.2 keV of each location are shown. The ground truth of the interaction positions was extracted from the setup file of the GATE simulation.

### 2.3 Evaluating the spatial resolution and bias

Using the testing dataset, the detector's spatial resolution was evaluated by comparing the predicted position, '*x(predicted)*' and '*y(predicted)*' with the ground truth, '*x(true)*' and '*y(true)*'. Histograms of the differences between 'predicted' and 'true' values along X- and Y-directions were plotted in Figure 6 and fitted with a Gaussian function. The FWHM (full width at half maximum) from the fit provides an estimate of the spatial resolution of the detector. The centroid of the fit gives an estimation of the average bias for the predicted value from the ground truth value.

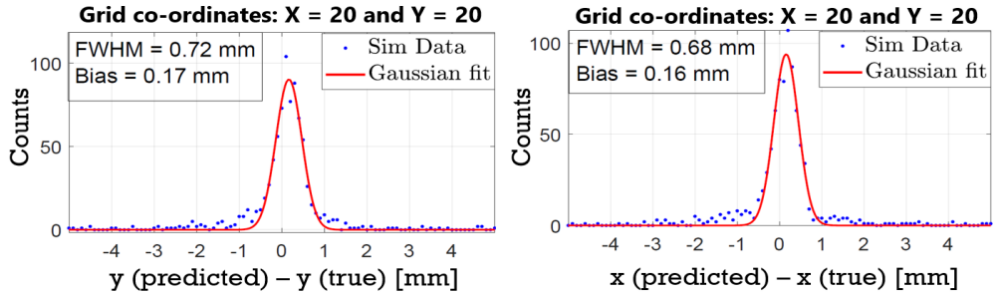

*Figure 6: FWHM results for one testing grid point with X = 20 and Y = 20, located towards the center of the X-Y plane for the case of 32 signal readout w.r.t. (a) Y-direction and (b) X-direction.*

When the absolute value of the bias is smaller than the pitch between two neighboring test points, bias correction can be effectively applied in the predicted value, following Equation 2.

$$x(predicted\ after\ bias\ correction) = x(predicted) - x(bias), \qquad (2)$$

where '*y (predicted after bias correction)*' was also obtained following Equation 2.

After correction of bias from the predicted value, histograms of the differences between 'bias corrected value' and 'true' values can be plotted along X- and Y- directions (similar to Figure 6) to obtain the bias corrected spatial resolution. This method of bias correction was applied and subsequently presented in a discussion of the results.

Figure 7 summarizes the steps followed from generation of the training-test dataset to the evaluation of the spatial resolution in sequential order in the form of a flow chart.

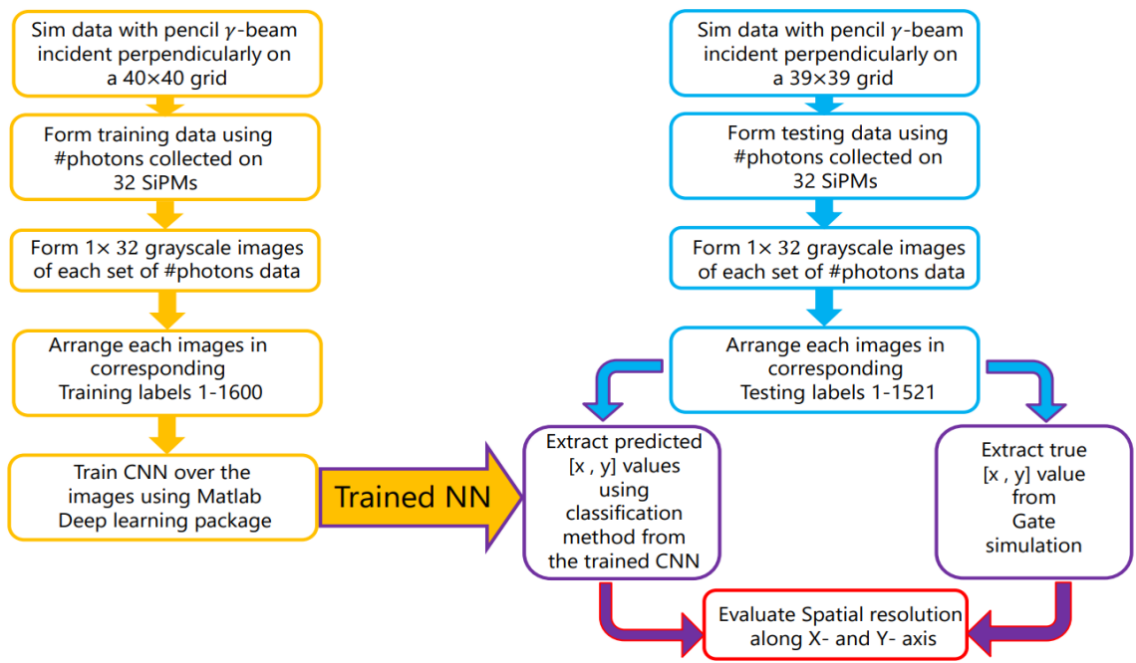

*Figure 7: Flowchart showing the various steps involved for evaluating the spatial resolution.*

*2.4 Testing multiplexing schemes*

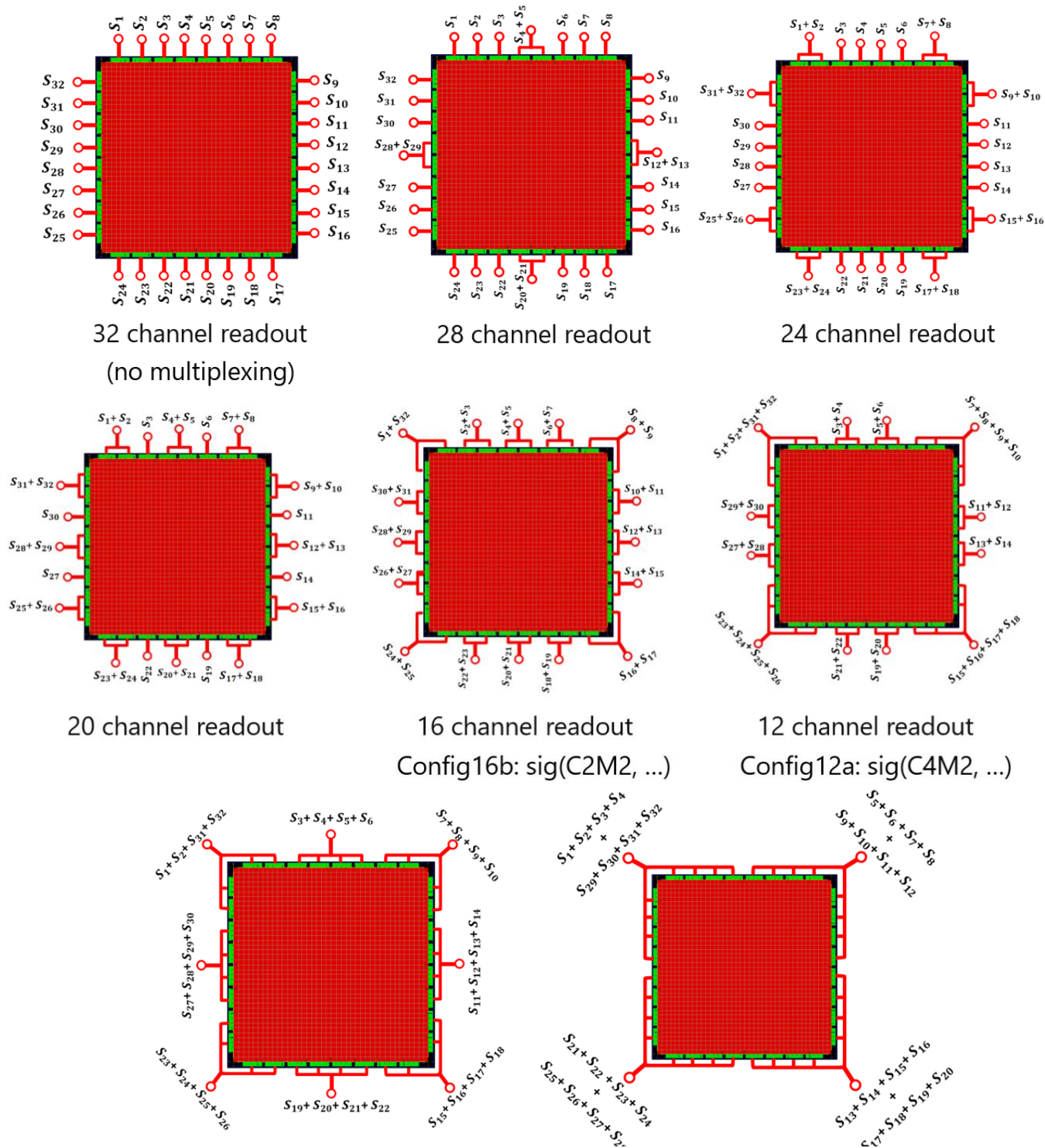

*Figure 8: Multiplexing schemes tested for readout channels 28, 24, 20, 16, 12, 8, and 4 each. For readout channels 28, 24, and 20, figure shows the only multiplexing schemes studied in this work and for readout channels 16, 12, 8, and 4, the figure shows the multiplexing schemes found to have optimal detector performance.*

After the scintillation light information is collected from each SiPM, various multiplexing schemes for combinations (i.e., summing) of the analog signals from the SiPMs before digitization were tested (Figure 8). The multiplexing schemes were grouped based on the final number of readout channels after summing the analog signals. The readout channels were decreased by 4 starting with the 32-channel readout until total readout channels reached 4. For some readout channels, e.g., 28, 24, and 20, only one multiplexing scheme was studied and for other readout channels, e.g., 16, 12, 8, and 4, more than one multiplexing scheme were studied. However, as shown in Figure 8, only a single multiplexing scheme (the one which provided best performance) for each number of readout channels is presented. The other schemes tested have been included in the '*Supplementary Information*'.

## 3. RESULTS

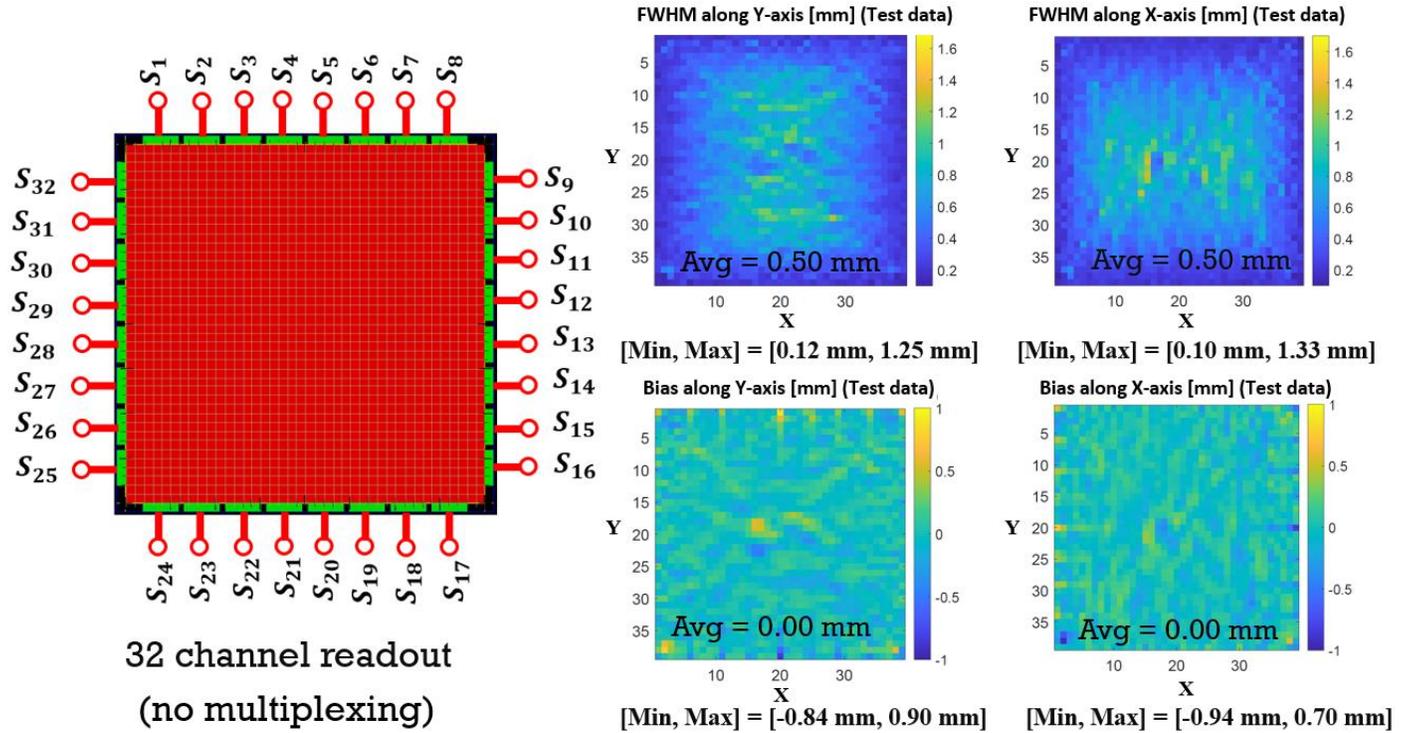

*Figure 9: Spatial resolution and bias results for the case of signal collection from each of 32 SiPMs (no multiplexing).*

The 2D histograms of FWHM and bias results for the case with no multiplexing is shown in Figure 9. Signals from the 32 SiPMs are labeled as $S_1$, $S_2$, ..., $S_{32}$. The corresponding FWHM and bias histograms, their average (Avg) and the minimum and maximum values ([Min, Max]) are also given. On average, submillimeter (~0.50 mm) FWHM along both X- and Y-direction was obtained. From this analysis, it was found that in general, the value of FWHM is better at the corner and edges of the crystal surface (0.10 mm along X-direction and ~ 0.12 mm along Y-direction and) compared to the central region (~1.33 mm along X-direction and ~ 1.25 mm along Y-direction and). The reason for the relatively poor spatial resolution in the central region of the detector plane compared to the corner and edge region is likely due to the fact that for a gamma-ray event closer to the edge and corner, the light distribution changes more rapidly and the neural network finds it easier to train over the feature maps of the grayscale images of these regions compared to the central region. The magnitude of bias was small (< 1 mm) over the entire detector plane, and in general it was found to increase in the corner regions and along the edge of the detector plane which were not covered by the SiPM (as seen in the detector geometry along with test data points in Figure 4). This result is consistent with similar published work which observed increase in bias as gamma-ray events approached the crystal edges where photon sensors were missing by design [2] [23] [24] [25]. The pitch in between the two test points is 1 mm and as observed, the bias didn't exceed 1 mm even in the worst-case scenario. This suggests that for the case of no-multiplexing, the positioning bias can be corrected with no ambiguity.

Below, the results of the various multiplexing schemes that were implemented for the readout channels: 28, 24, 20, 16, 12, 8, and 4 are discussed.

### 3.1 Impact of multiplexing schemes on the light distribution of input grayscale images to NN

As the analog signals from 32 SiPMs are combined with different multiplexing schemes, the resultant grayscale images will have varied light distribution patterns as shown in Figure 10. In the figure, three sample gamma-ray incident events were chosen at the middle (M), edge (E), and corner (C) regions of the detector plane (as shown in Figure 2) and the effect of multiplexing (from Figure 8) on the corresponding grayscale images is shown as the readout channels was decreased from 32 to 4 with a readout reduction interval of 4.

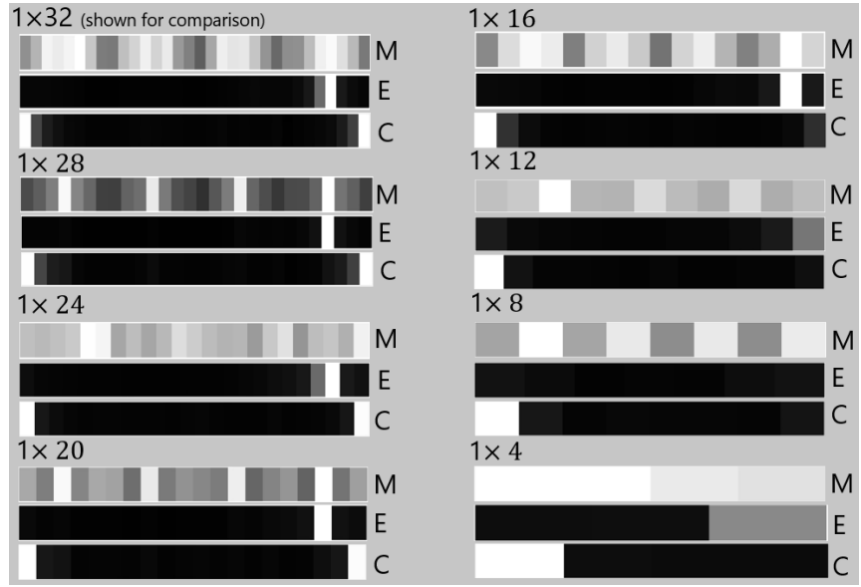

*Figure 10: Scintillation light from 32 SiPMs being converted into grayscale images for the multiplexing schemes shared in Figure 8. 'M', 'E', and 'C' represent the three-interaction location of gamma-rays at the middle, edge, and corner regions of the detector plane, as shown in Figure 2.*

From Figure 10, the following three observations can be made:

1. Across all different number of readout channels, for events near the corner and edge regions as compared to the middle region, since almost all the 32 SiPMs (with exception to the SiPM(s) close to the gamma-interaction location for edge and corner regions) receive low photon statistics, the light variation in between 32 pixels in the corresponding grayscale images is small.
2. As the number of readout channels is decreased, the events near the edge and corner regions can have similar patterns of light distribution.
3. As the number of readout channels is decreased, the dimension of the input image also decreases, which will reduce the number of feature maps of the images for CNN to train on (e.g., 32 feature maps for 32-channel readout vs 4 feature maps for 4-channel readout).

### 3.2 Multiplexing study results

Following the same approach discussed for the no-multiplexing scheme, the FWHM and bias values were extracted for all multiplexing schemes (from Figure 8) over the 39 × 39 grid testing positions and the corresponding 2D histogram color maps are shown in Figure 11.

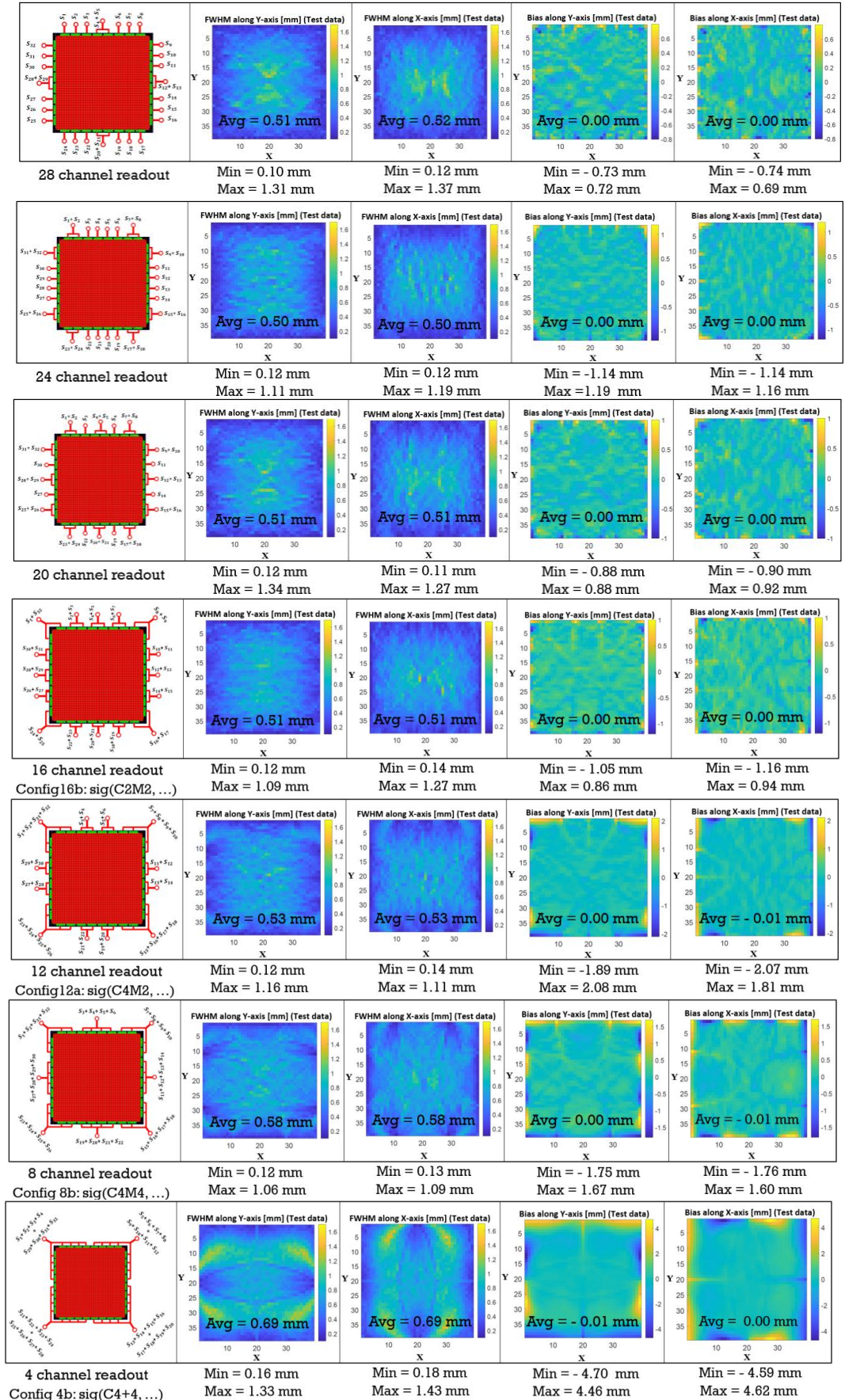

*Figure 11: FWHM and bias results obtained for the corresponding multiplexing schemes shared in Figure 8. For all the readout channels, the figure shows the average (Avg) of the FWHM and bias from each of 39 × 39 test grid along with their respective minimum (Min) and maximum (Max) values.*

From Figure 11, along with the remaining multiplexing schemes shown in the '*Supplementary Information*', across most of the readout channels, the following three observations can be made:

1. FWHM is generally better at the edge and corner regions of the detector plane and poor towards the central region. A logical basis for this observed pattern can be inferred from the light distribution of the grayscale images fed into the CNN. From Figure 10, it can be seen that for events near the corner and edge regions as compared to the middle region, there is minimum variation of light distribution in 32 pixels, so the CNN is training over less complex features of the input images. This likely made it easier for the CNN to identify events in proximity to these regions (i.e., CNN is able to assign high probability for the interaction location to some nearby locations for events in these regions).

2. As the number of readout channels is reduced, bias is generally better in the central region of the detector plane and poorer towards the edge and corner regions. A likely explanation for this can again be inferred from the light distribution of events for these two regions. From Figure 10, it can be seen that, the events near the edge and corner regions can have similar patterns of light distribution as the number of readout channels is decreased, which can make it difficult for the CNN to identify events at the corner vs edge regions. In addition, there is no coverage of SiPMs at the very 4 corners of the detector volume (1.9 mm on each corner from the vertex). This could also have some contribution to high bias at the corner regions of the detector plane.

3. The FWHM is observed to improve in the central region as the number of readout channels is decreased. The likely reason behind this can also be linked to the light distribution pattern from Figure 10. From the light distribution, it can be observed that as the readout channels is decreased, this will reduce the feature maps of the images for CNN to train. Improved training on relatively fewer features leads to the FWHM improving with a reduction of the number of readout channels.

### 3.3 Summary of the multiplexing study results

For convenience of comparison of the FWHM results over the different multiplexing schemes tested above, in Figure 12, a scatter plot is drawn below for average FWHM values along X- and Y-directions of the detector plane. Since the distribution of SiPMs is symmetric along X and Y direction, the average FWHM values is expected to be similar along the X-Y direction.

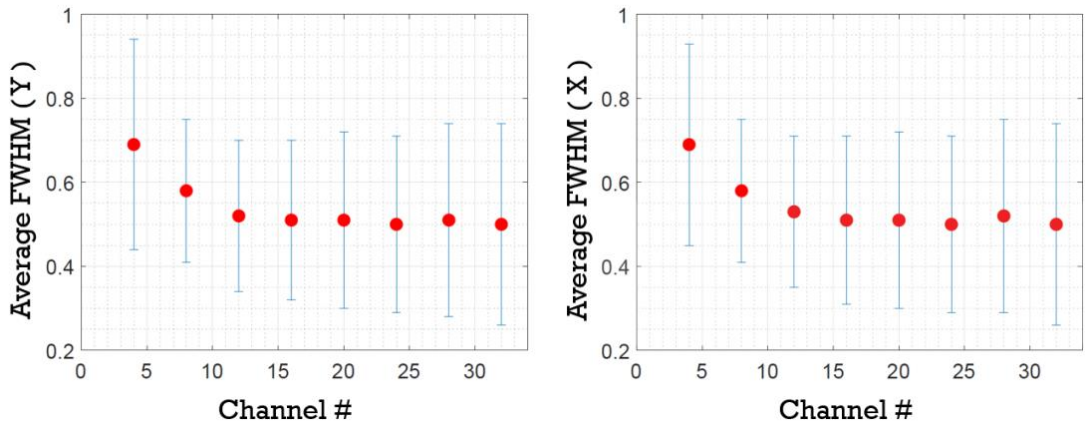

*Figure 12: Scatter plots of average FWHM as a function of channel # for (left) Y-direction and (right) X-direction for the multiplexing schemes shared in Figure 8. The error bars represent the standard deviation of the 1521 FWHMs from the testing data set.*

As shown by the FWHM of the 2D histograms in Figure 11, FWHM values showed some variability across the $39 \times 39$ grid testing positions bounded by the respective [Min, Max] values. The standard deviation ($\sigma$) of the FWHM was calculated for all the multiplexing schemes across all readout channels to quantify the variability of FWHM values across the $39 \times 39$ grid testing positions with respect to its average value. The error bar drawn represents the standard deviation of individual FWHM values over the $39 \times 39$ grid testing positions w.r.t. the average FWHM.

From Figure 12, the following observations can be made:

a) The average value of the FWHM remains mostly constant when the number of readout channels is lowered from 32 to 16, however any further reduction of readout channels would lead to degradation in the FWHM.

b)  As observed in Figure 12, an interesting behavior was observed for the $\sigma$, i.e., although $\sigma$ is largest for the lowest number of readout channels, as the number of readout channels is increased beyond 16, it doesn't necessarily translate to smaller $\sigma$.

### 3.4 Impact of multiplexing on bias

From the 2D histogram plots of bias for different readout channels, it was found that bias is highest in the corner and edge regions of the detector module compared to the central region. Bias was observed to be least for the 32-channel readout (no-multiplexing) and highest for the 4-channel readout. For the 32-channel readout, in order to investigate the impact on bias, a grid of 3×3 test positions were chosen across all four corners of the detector module where the bias is observed to be greater and the histograms of '*x/y(predicted) – x/y(true)*' values were plotted for all the four-corner 3×3 test positions, as shown in Figure 13. As seen in the figure, a single total histogram for the corner points is also drawn by including only the data from the above four corner 3×3 test positions. From the individual data points in the total histogram for the corner points, the mean squared error (MSE) value was computed, which gives a measure of the bias effect. For the 32-channel readout (Figure 13), it was observed that the effect of bias extends to only within the nearest neighboring test grid point i.e., < 1.0 mm. This causes the individual histograms drawn to be clustered within 1 mm on either side of the '*true*' position and the total histogram to peak within the range.

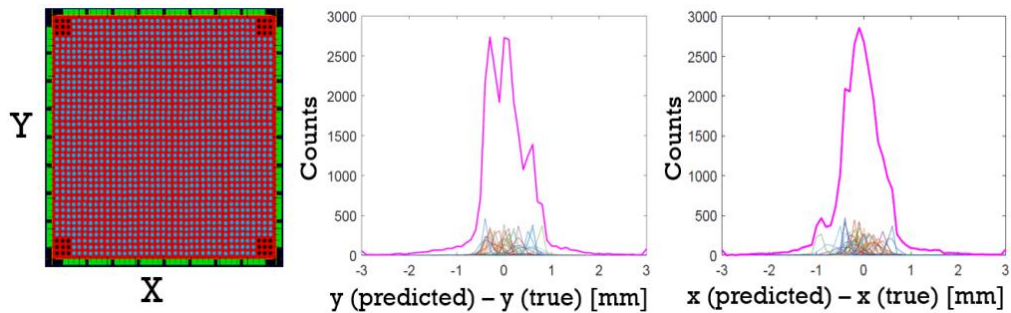

*Figure 13: Comparison of bias for 32 channel readout with 3×3 corner grid test points (marked with 'black') on all four corners of the detector plane. Individual histograms represent the CNN prediction of interaction location for test images for each 3×3 corner test points. The data from each individual histogram are then combined to form a total single histogram which shows the spread of bias in the corner regions.*

A similar process was also performed for the 4-channel readout for a larger grid of 5×5 test positions, as shown in Figure 14 (to reflect the increased region with bigger bias). For the 4-channel readout, as seen in the figure, the impact on bias was observed to be much more significant (extending as far as four neighboring test grid points i.e., ~ 4.0 mm on either side of the '*true*' position). For these corner test grid points, though NN is able to predict the interaction location within nearest neighboring test grid points for some of the test sample points (causing corresponding individual histograms to be clustered within [-1, 1] mm), the prediction from NN was observed to extend as far as four neighboring test grid points. Since the NN predicts the probability of interaction location based on the training grid, it causes the individual histograms to be spaced and peak at several nearby locations corresponding to training grid points, as shown in Figure 14.

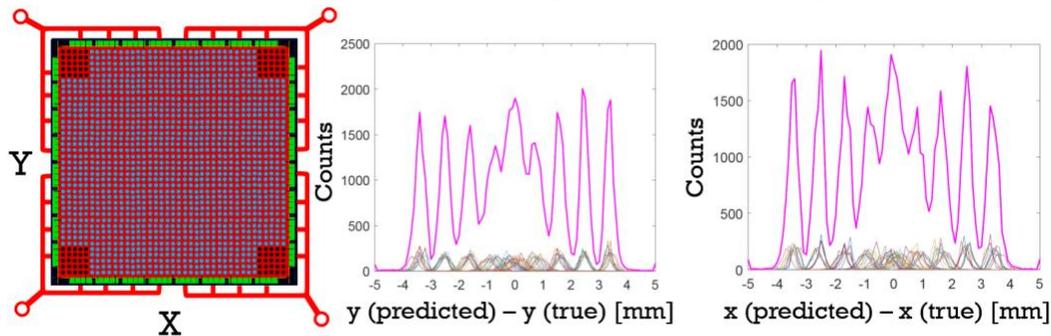

*Figure 14: Same as Figure 13, but for the 4-channel readout with 5×5 corner grid test points. Here, compared to Figure 13 the impact of bias is observed to be more significant.*

An analysis of the MSE was performed for the optimal configurations across all channel readouts and the results are shown in Table 1. As seen from the table, similar to the average FWHM values w.r.t. channel # (Figure 12), the MSE also remains fairly constant when the number of readout channels is decreased from 32 to 16, however any further reduction of readout channels leads to an increase in the MSE. This implies, the impact on bias remains nearly constant when decreasing the readout channels from 32 to 16, and increases with a further decrease in the number of readout channels.

*Table 1: Mean Square Error (MSE) values calculated for the test data points at the corners of detector module for various channel numbers. Since, the effect of bias increases with decreasing numbers of channels, the size of the corner grid of test data points is adapted accordingly. The multiplexing schemes are the ones shown in Figure 8.*

| Channel # | MSE_Y | MSE_X | Corner grid |
|-----------|-------|-------|-------------|
| 32 | 0.40 | 0.38 | 3 × 3 |
| 28 | 0.41 | 0.39 | 3 × 3 |
| 24 | 0.61 | 0.60 | 4 × 4 |
| 20 | 0.56 | 0.53 | 4 × 4 |
| 16 | 0.48 | 0.49 | 4 × 4 |
| 12 | 1.18 | 1.09 | 5 × 5 |
| 8 | 0.99 | 0.97 | 5 × 5 |
| 4 | 4.33 | 4.51 | 5 × 5 |

### 3.5 Impact of bias on spatial resolution and test of bias correction

In this section, the impact of bias on neighboring test grids is discussed based on the highest (32) channel and least (4) channel readout schemes. The objective behind this investigation is based on the idea that if the bias has value larger than the test grid pitch size (1 mm), then this will negatively impact the confidence on the spatial resolution (FWHM) around neighboring test grid points, i.e., for some test grid points near the high bias region at the corners of the detector plane (more pronounced for low channel readout configurations), NN consistently misclassifies the predicted value away from the true value. Though this falsely seems to improve the spatial resolution (higher precision in prediction), it reduces the accuracy of the prediction and the spatial resolution.

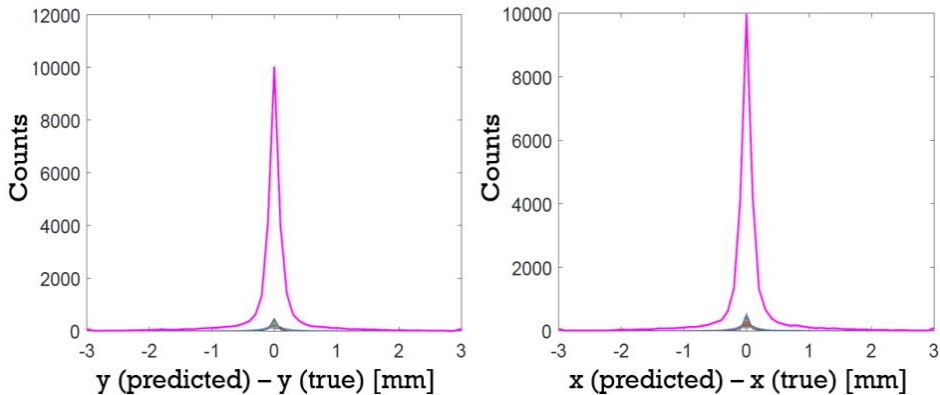

*Figure 15: Applying bias correction to the corner test grid points for the 32-channel readout scheme. Here, bias correction is firstly applied to the predicted values (following Equation 2) and then the histograms of their difference with true values is plotted. As observed, after bias correction the individual histograms for each corner test grid point are now centered around the x(predicted) = x(true) position. The total histogram maintains a gaussian peak shape.*

For instance, as seen in Figure 13, for the 32-channel readout scheme, the impact of bias is limited within the test grid pitch size of 1 mm, even in the highest bias region. As the impact of bias on neighboring test grid points is not observed, bias correction can too ideally be applied for the 32-readout scheme for each of the corner test data points. On applying the bias correction, all the individual histograms for each corner test data points get centered around *0* and the total histogram retains a gaussian peak shape, as shown in Figure 15. The total histogram was fitted with a double gaussian function and the bias corrected FWHM was obtained for the region, as shown in Figure 16.

However, as observed from Figure 14, for the 4-channel readout case, the impact of bias is observed across several neighboring corner test grid points. This degrades the confidence on the evaluated FWHM in those high bias regions and thus the bias correction cannot be applied. The MSE in Table 1, provides a basis for comparison of the influence of bias on the detector's spatial resolution for all other readout schemes when compared with the 32- and 4-channel readout schemes.

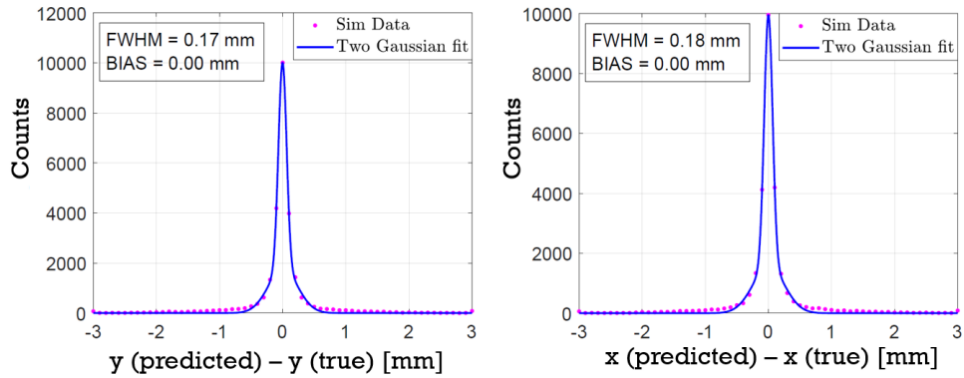

*Figure 16: The total histogram for the 32-channel readout scheme from figure 15 is fitted with a double gaussian function and the corresponding FWHM is evaluated for the X and Y directions.*

### 3.6 Applying bias correction on the spatial resolution for entire grid of 32-channel readout

From Figure 16, it can be observed that as the bias correction is applied, FWHM improves and is close to the best FWHM results that was obtained for the 32-channel readout (Figure 9). This shows that the spatial resolution can be recovered from the impact of bias even from the high bias region if the bias is limited within the pitch size of the 1 mm grid used for training/testing. Bias correction was implemented for the entire $39 \times 39$ grid of test positions and the corresponding histograms of '$x/y(predicted) - x/y(true)$' values are shown in Figure 17. The histograms were fitted with a double Gaussian function to evaluate the FWHM and bias values. As observed from Figure 17, with bias correction FWHM = 0.34 mm (both X- and Y-direction) for the entire detector plane, which is significantly better ($> 2\sigma$) than the average FWHM = 0.50 mm (both X- and Y-direction) obtained previously from the bias uncorrected 2D histogram color maps (Figure 9). For adding additional perspective in evaluating the spatial resolution, Figure 17 also includes the full width at tenth maximum (FWTM) value = 1.12 mm (both X- and Y-direction).

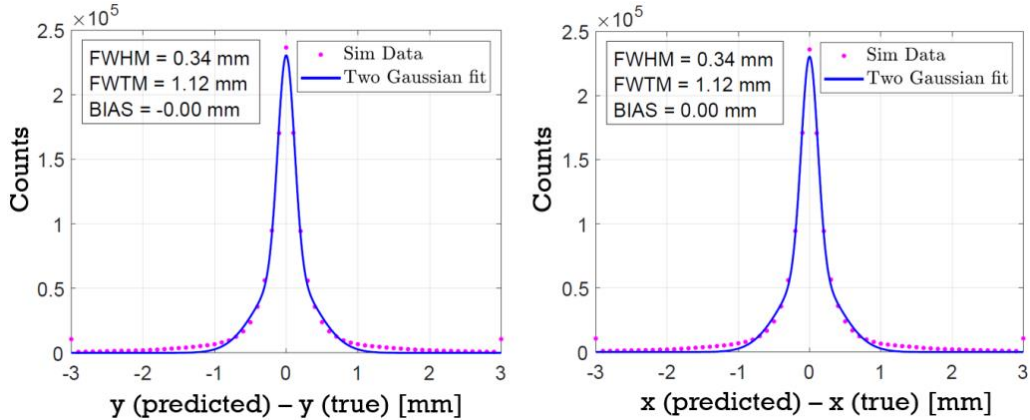

*Figure 17: Applying bias correction for the entire $39 \times 39$ test grid for the 32-channel readout. After implementing a double gaussian fit, corresponding values of FWHM, full width at tenth maximum (FWTM), and bias are evaluated along X- and Y-directions of the detector.*

## 4. DISCUSSION

This work utilizes machine learning (CNN architecture) to evaluate the spatial resolution of a semi-monolithic detector module proposed by our group [4] [5]. One of the goals of this study was to perform extensive research on ways to improve the past CNN architecture for optimal performance. Using the common Neural Network Toolbox in MATLAB, different variations in CNN hyperparameters and training options were tested out and an optimal CNN architecture is presented from among the options tested. Although the best of the CNN architectures found to date led to better overall training and achieved improved spatial resolution, there are still regions of poor spatial resolution (middle region of the detector plane) and high bias (edges and corners of the detector plane) which could be an intrinsic outcome of the detector design and CNN architecture for positioning events that is difficult to improve upon. Similar to this work, [26] decoded the scintillation location from multiple monolithic LYSO crystals based on the light patterns measured by SiPMs and used a variety of more advanced CNN architectures (e.g., ConvNeXt, ResNet, and EfficientNet families) to determine the interaction position. While implementing deep learning with neural networks in evaluating the interaction (scintillation) location of gamma-rays, in addition to the classification approach (as implemented in this work), there also has been investigation of using the regression approach. [27] using both a CNN architecture (classification approach) and a fully connected network (classification and regression approach) in predicting the interaction location in monolithic crystals. Similarly, [28] used a ten layered deep-residual (DR)-CNN architecture to decode the interaction location for their monolithic annular PET scanner. Variations in deep-learning approaches for event positioning represents a wide-open field for further exploration.

Though the bias is observed to be high at the corners and edges of the detector plane, we estimate that the overall impact on image quality is negligible even for the highly multiplexed schemes when a likelihood maximization-based reconstruction method is used and a proper probability distribution function is implemented. The probability distribution function representing the detector's positional response, i.e., the point spread function (PSF), shall include not only the resolution of the detector, but also the probability distribution for the bias. This will ensure that no image artifact would occur due to the bias at corners. In addition, the small fractions of events detected in the high bias region will introduce little degradation to the overall image quality.

There are several limitations to the current work. The incident 511 keV gamma-rays can scatter in between different crystals via Compton interaction before being photo-electrically absorbed. This work is only based on a single layer monolithic detector module, and so the effect of inter-crystal scattering was not considered. To build a realistic PET system, a series of thin monolithic scintillator plates need to be stacked forming a semi monolithic detector design, in which case the problem of inter crystal scattering will become significant. In order to understand the impact of inter-crystal scattering, a similar study needs to be performed with a multi-layered detector module. However, for the multi-layered detector, implementing the machine learning technique (as implemented here for a single layer detector) to predict the interaction location will have some challenges. The ML positioning algorithm will need to be able to predict the first interaction location of the gamma-rays in crystal by learning to distinguish between Compton scatter interaction and photoelectric interactions. It would still be possible to apply the current CNN architecture for a multi-layered detector, but for this, the following two aspects should be considered: (a) in addition to the ground truth interaction location information for the gamma-rays, the energy deposited by the gamma-rays for each interaction should be recorded and (b) a lot more data should be generated to train a network in order to cover the whole phase space of Compton scattering probabilities.

The optical model used in this work doesn't investigate the effects of various crystal and SiPM properties. However, since the main purpose of this work is to test and compare the effect of various multiplexing schemes, the results presented in this paper shouldn't be tied with the above limitation.

In the current work, detector performance was evaluated based on the gamma-interaction with the detector corresponding to interleaved training and test datasets ($40 \times 40$ grid of training and $39 \times 39$ grid of test data points). However, a gamma-interaction can happen at any location in the detector plane. Thus, the current setup, where the test dataset is as different as possible from the training dataset, arguably presents the worst-case scenario. A more realistic test case could consider interactions at some random position in the detector plane. This could potentially lead to relatively better detector performance than currently found. However, the scenario of realistic random gamma-interactions would present an impractical difficulty in gathering sufficient statistics, as it would necessitate a repetition of gamma-ray incidents at the same random position to assess the FWHM and bias at that particular location.

Another limitation of this study is that electronic noise is not modeled, which could have a small impact in the detector's performance as signals are multiplexed together. However, it is expected that statistical fluctuations in the photon distribution are the dominant source of uncertainty here. Future experimental investigations will verify this as a single layer monolithic PET detector module has now been built and its experimental performance, guided by the different multiplexing schemes evaluated in this work, will be a key next step in further developing this detector concept.

## 5. CONCLUSION

From the multiplexing schemes tested in this work for various numbers of readout channels, it was found that as the number of readout channels is reduced from 32 to 16, the average FWHM remains fairly constant. This shows that there is certain room to reduce the number of readout channels without sacrificing the detector's spatial resolution. The motivation to reduce the number of readout channels comes from the fact that analog signals from SiPMs can easily be combined before digitization via various multiplexing schemes, which will help reduce electronics cost and power consumption.

From the multiplexing tests, the following observations about the spatial resolution (FWHM) and bias results can be derived. Spatial resolution is generally better when the scintillation event occurs close to the edge and corner regions of the detector module compared to the center region. The bias is negligible when the number of readout channels is decreased from 32 to 16. However, when the number of readout channels is reduced below 16, the bias in corner areas of the detector module starts to grow. For the case where the number of readout channels equals 12, bias becomes non-negligible and for the case when readout channels are reduced to 4, bias becomes most prominent.

This sample of multiplexing schemes tested is only a subset of multiplexing schemes that could be tried and tested out. The goal of this work was to investigate if optimal detector performance can be obtained even when the number of readout channels is reduced by suitable multiplexing scheme, which is clearly observed to be the case. Compared to all other channel readout schemes, considering the average FWHM and corresponding $\sigma$, 16-channel readout seems to be the lowest number of channels that can preserve performance.

As demonstrated in the multiplexing schemes shared in this work, for different number of readout channels, there can be various ways in which signals from the SiPMs can be combined while maintaining a constant number of readout channels. In this investigation also, it was found that in certain cases, combining signals from SiPMs located at the corners of two neighboring sides could potentially lead to relatively better performance in FWHM and bias.

## 6. ACKNOWLEDGEMENTS

This work was funded by NIH grant R01EB028337. The authors would like to thank collaborators from the Department of Biomedical Engineering at UC Davis for valuable discussions.

# 7. SUPPLEMENTARY INFORMATION

This section includes a discussion of all the multiplexing schemes for readout channels 16, 12, 8, and 4 that were evaluated (excludes multiplexing schemes for 28, 24, 20 channel readout that were presented in the main manuscript along with the 32-channel readout).

### 7.1 Discussion of the additional multiplexing schemes

#### 7.1.1 16-channel readout

Figure 18 shows the total 7 different multiplexing schemes that were investigated for the 16-channel readout scheme (including the optimal scheme). The 7 different multiplexing schemes are labeled as: Config16a, Config16b, …, Config16g. The FWHM and bias results obtained for each of the configurations are presented in Figure 18. It was found that for the 7 configurations investigated, the best average FWHM was 0.51 mm (for Config16a and Config16b) while the worst average FWHM was 0.70 mm (for Config16c and Config16d). These results indicate that even for the case when the number of readout channels is the same, not all multiplexing schemes provide similar performance, and thus highlights the significance of this multiplexing study. Comparing the two configurations with the best FWHM results i.e., Config16a and Config16b, when bias results are taken into consideration, it was found that Config16b has better bounds in the magnitude of the bias compared to Config 16a. This result highlights the case that in certain multiplexing schemes, configuration where the signals from SiPMs at the corner of the detector module are combined can provide relatively better performance.

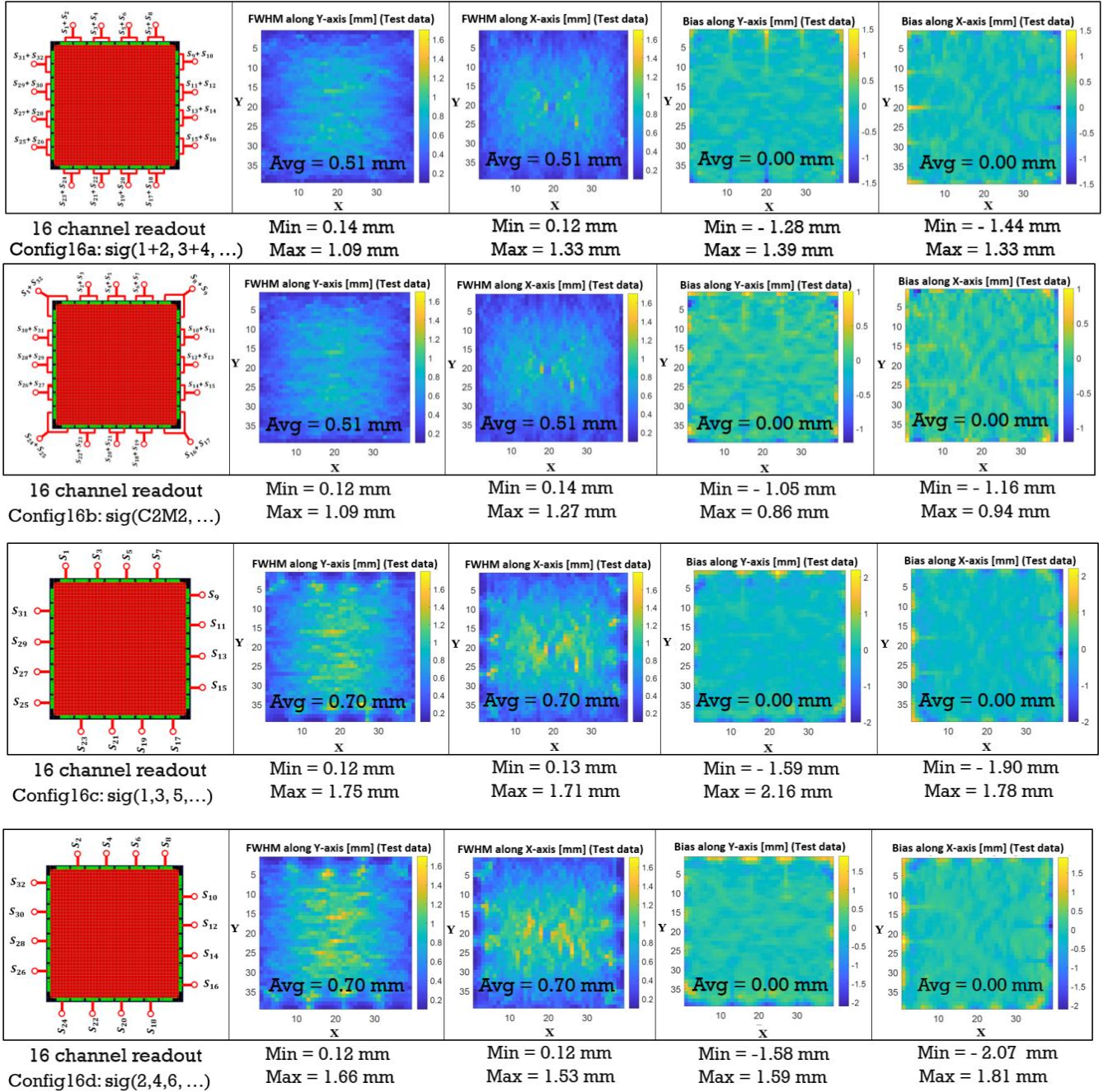

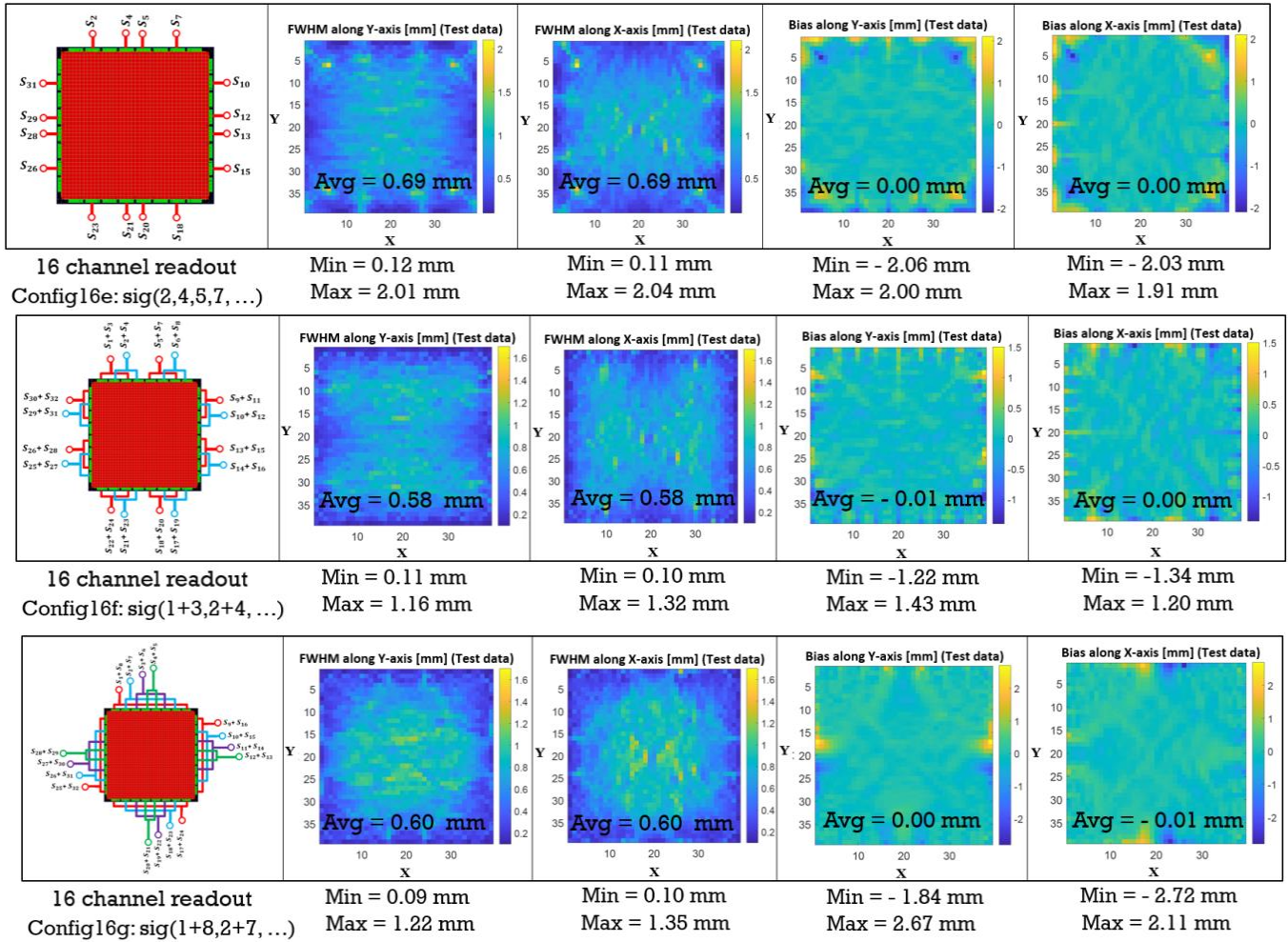

*Figure 18: Multiplexing schemes investigated for the 16-channel readout. Here, we present the results for 8 different multiplexing schemes we investigated.*

### 7.1.2 12-channel readout

In Figure 19, the 4 different multiplexing schemes (including the optimal scheme) investigated for the 12-channel readout scheme are presented. The 4 different multiplexing schemes are labeled Config12a, …, Config12d. The FWHM and bias results showed that all the configurations tested provided similar performance. However, the best average FWHM results were obtained for the Config12a configuration and the worst FWHM results were obtained for the Config12b configuration. The magnitude in upper bounds of the bias exceeded (~ 1 mm) for all the schemes tested, especially in the corner regions and edge regions. It was also noticed that the bias in some configurations (Config12b and Config12d) has relatively smaller bounds in magnitude compared to the other two configurations. This suggests that there is room to explore new variations in multiplexing for optimal FWHM and bias performance.

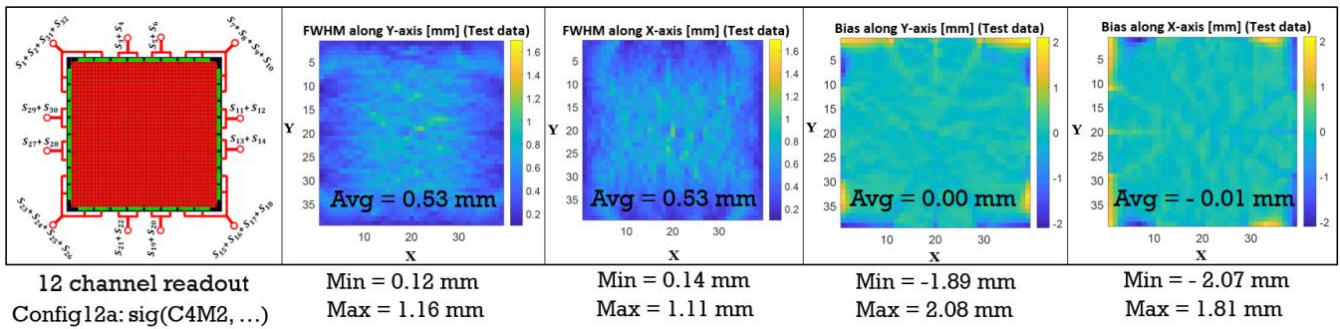

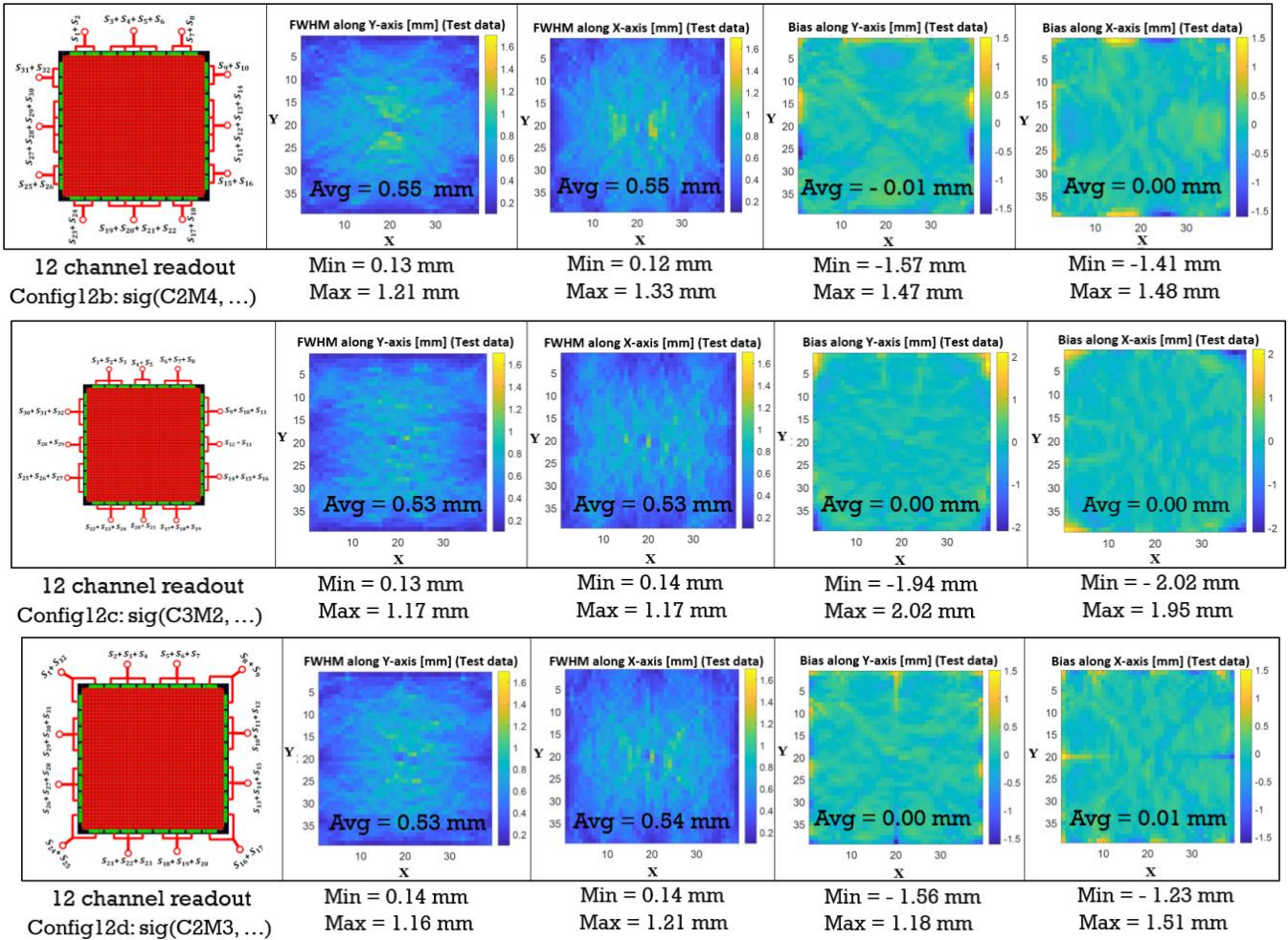

*Figure 19: Multiplexing schemes investigated for the 12-channel readout along with their FWHM and bias results.*

### 7.1.3 8-channel readout

For the 8-channel readout scheme, 2 different multiplexing schemes were investigated, which are labelled Config8a and Config8b, as shown in Figure 20. Considering the FWHM results, both schemes provided comparable results. However, when considering the bias results, for the upper limit of bias (in magnitude) in the corner and edge regions of the detector module, Config8b provided relatively better results compared to Config8a. This result further supports the observation from 16-channel and 12-channel readout and highlights the case that in certain multiplexing schemes, configuration where the signals from SiPMs at the corner of the detector module are combined can provide relatively better detector performance.

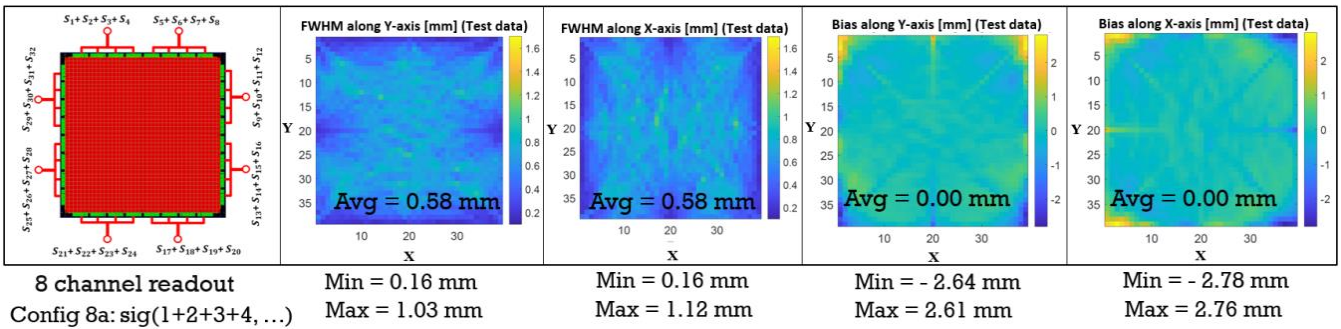

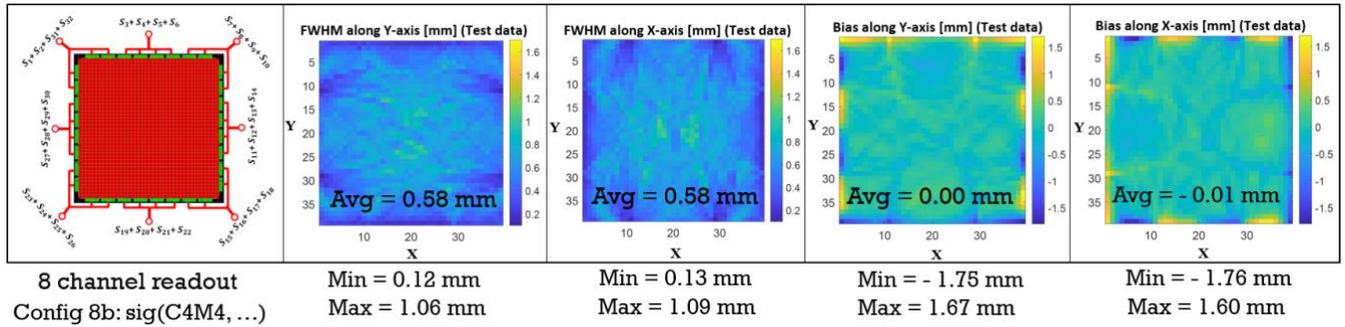

*Figure 20: Multiplexing scheme investigated for the 8-channel readout.*

### 7.1.4 4-channel readout

In the 4-channel readout configuration, two multiplexing schemes were investigated, which are labelled Config4a and Config4b, as shown in Figure 21. Comparing the two multiplexing schemes, it was found that Config4b provided relatively better average FWHM results. From the bias 2D histogram, it was found from the [Min, Max] bias values that Config4b provided relatively better results. Compared to the readout schemes with larger numbers of channels, the 4-channel readout has the most bias and this bias is non-negligible.

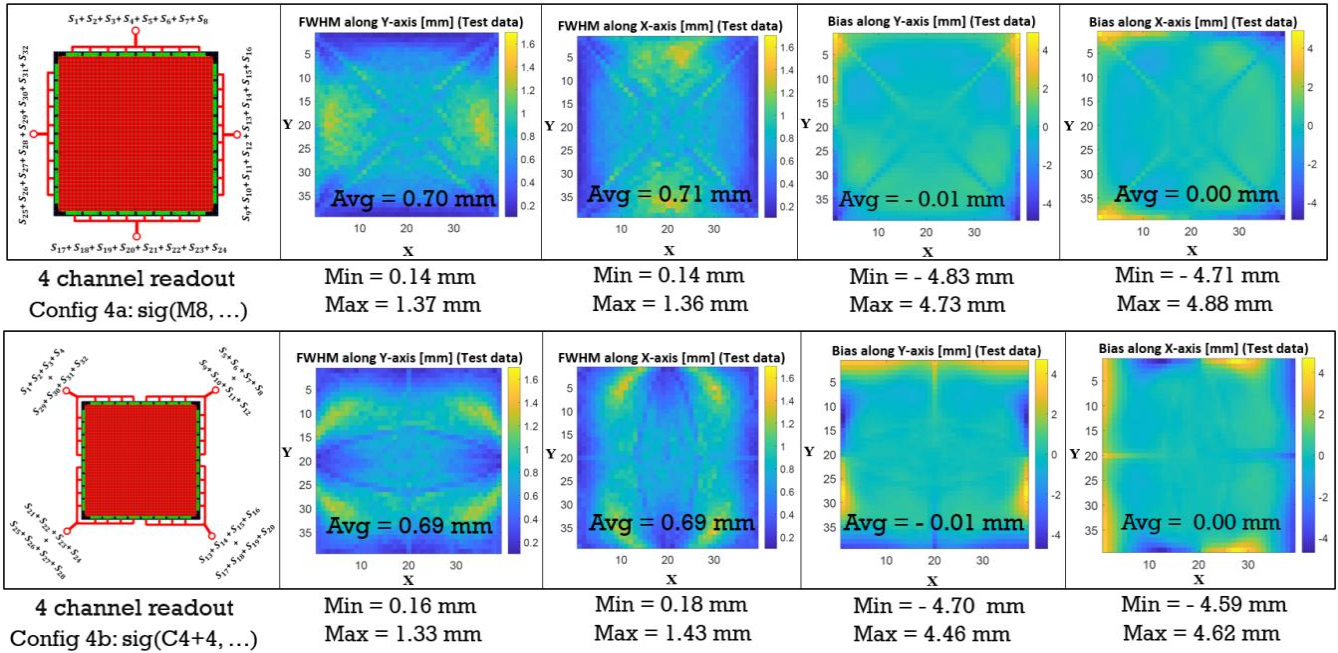

*Figure 21: Multiplexing scheme investigated for the 4-channel readout (excluding the optimal scheme).*

### 7.2 Summary of the results for all multiplexing schemes

In Table 2, the results of the average FWHM are presented for all the multiplexing schemes investigated w.r.t their readout channels along both the X- and Y- direction. In the table, the multiplexing configurations with optimal performance are highlighted in bold text for each degree of multiplexing. The standard deviation is also presented in Table 2 for all the multiplexing schemes across all readout channels along X- and Y- direction. In Figure 22, an extension of the scatter plot (Figure 12) is presented, which includes the FWHM results for all the configurations tested in this work. In Figure 22, for readouts with 28, 24, and 20 channels (with a single multiplexing scheme), the results are present in solid red circles along with 32 channel readout. For readouts with 16, 12, 8, and 4 channels, the multiplexing scheme with optimal performance have been shown in solid red circles while all other multiplexing schemes are shown for comparison purpose in hollow red circles.

Table 2: Summary of the average FWHM results along with the standard deviation (along X- and Y- direction) for the readout channels 32, 28, 24, 20, 16, 12, 8, and 4.

| Index | Channel # | Avg. FWHM (Y) | S.D. [$\sigma$] (Y) | Avg. FWHM (X) | S.D. [$\sigma$] (X) | Description |
|---|---|---|---|---|---|---|
| **1** | **32** | **0.50 mm** | **0.24 mm** | **0.50 mm** | **0.24 mm** | **32 pix 32 sig (no multiplexing)** |
| **2** | **28** | **0.51 mm** | **0.23 mm** | **0.52 mm** | **0.23 mm** | **32 pix 28 sig** |
| **3** | **24** | **0.50 mm** | **0.21 mm** | **0.50 mm** | **0.21 mm** | **32 pix 24 sig** |
| **4** | **20** | **0.51 mm** | **0.21 mm** | **0.51 mm** | **0.21 mm** | **32 pix 20 sig** |
| 5 | 16 | 0.51 mm | 0.20 mm | 0.51 mm | 0.20 mm | Config 16a: sig(1+2, 3+4, …) |
| **6** | **16** | **0.51 mm** | **0.19 mm** | **0.51 mm** | **0.20 mm** | **Config 16b: sig(C2M2, …)** |
| 7 | 16 | 0.70 mm | 0.27 mm | 0.70 mm | 0.28 mm | Config 16c: sig(1, 3, 5, …) |
| 8 | 16 | 0.70 mm | 0.27 mm | 0.70 mm | 0.27 mm | Config 16d: sig(2, 4, 6, …) |
| 9 | 16 | 0.69 mm | 0.27 mm | 0.69 mm | 0.27 mm | Config 16e: sig(2, 4, 5, 7, …) |
| 10 | 16 | 0.58 mm | 0.22 mm | 0.58 mm | 0.22 mm | Config 16f: sig(1+3, 2+4, …) |
| 11 | 16 | 0.60 mm | 0.26 mm | 0.60 mm | 0.26 mm | Config 16g: sig(1+8, 2+7, …) |
| **12** | **12** | **0.52 mm** | **0.18 mm** | **0.53 mm** | **0.18 mm** | **Config 12a: sig(C4M2, …)** |
| 13 | 12 | 0.55 mm | 0.20 mm | 0.55 mm | 0.20 mm | Config 12b: sig(C2M4, …) |
| 14 | 12 | 0.53 mm | 0.18 mm | 0.53 mm | 0.18 mm | Config 12c: sig(C3M2, …) |
| 15 | 12 | 0.53 mm | 0.18 mm | 0.54 mm | 0.18 mm | Config 12d: sig(C2M3, …) |
| 16 | 8 | 0.58 mm | 0.17 mm | 0.58 mm | 0.18 mm | Config 8a: sig(1+2+3+4, …) |
| **17** | **8** | **0.58 mm** | **0.17 mm** | **0.58 mm** | **0.17 mm** | **Config 8b: sig(C4M4, …)** |
| 18 | 4 | 0.70 mm | 0.27 mm | 0.71 mm | 0.27 mm | Config 4a: sig(M8, …) |
| **19** | **4** | **0.69 mm** | **0.25 mm** | **0.69 mm** | **0.24 mm** | **Config 4b: sig(C4+4, …)** |

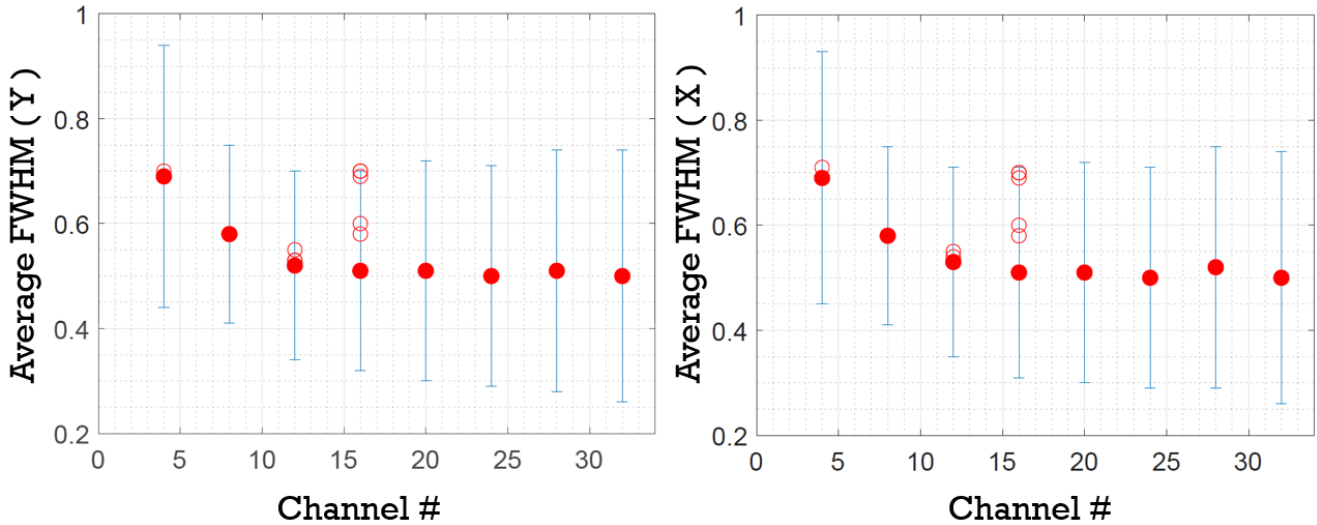

Figure 22: Scatter plots of average FWHM as a function of channel # for (left) Y-direction and (right) X-direction. The error bars represent the standard deviation of the 1521 FWHMs from the testing data set.